\documentclass[aps,preprintnumbers,floats, prd, twocolumn, nofootinbib]{revtex4}
\usepackage{}
\usepackage{amssymb}

\usepackage{graphicx}
\usepackage{dcolumn}
\usepackage{bm}
\usepackage{amsmath}
\usepackage{amsthm}
\usepackage{multirow}
\usepackage{enumerate}

\usepackage{amsfonts}
\usepackage{euscript}
\usepackage{ifthen}
\usepackage{psfrag}
\usepackage{slashed}
\usepackage{hyperref}
\usepackage{gensymb}
\usepackage{color}

\newcommand{\be}{\begin{equation}}
\newcommand{\ee}{\end{equation}}
\newcommand{\bea}{\begin{eqnarray}}
\newcommand{\eea}{\end{eqnarray}}

\newcommand{\vev}[1]{\left<{#1}\right>}
\newcommand{\bra}[1]{\left|{#1}\right>}

\begin{document}

\title{The Minimal UV-induced Effective QCD Axion Theory}

\author{Yu Gao$^{1}$}
\email{gaoyu@ihep.ac.cn}
\author{Tianjun Li$^{2,3}$}
\email{tli@itp.ac.cn}
\author{Qiaoli Yang$^{4}$}
\email{qiaoli\_yang@hotmail.com}

\affiliation{
$^{1}$~Key Laboratory of Particle Astrophysics, Institute of High Energy Physics,\\
Chinese Academy of Sciences, Beijing 100049, China\\
$^{2}$~Key Laboratory of Theoretical Physics, Institute of Theoretical Physics, Chinese Academy of Sciences, Beijing, 100190, China \\
$^{3}$ School of Physical Sciences, University of Chinese Academy of Sciences, Beijing, 100049, China\\
$^{4}$Department of Physics and Siyuan Laboratory, Jinan University, Guangzhou 510632, China}

\begin{abstract}
The characteristic axion couplings could be generated via effective couplings between the Standard Model (SM) fermions to a pseudo-Goldstone from a high-scale $U(1)$ Peccei-Quinn (PQ) symmetry breaking. Assuming that the UV-induced effective operators generate necessary couplings before the PQ symmetry breaking, and any low-scale couplings to the SM are restricted to the Yukawa sector, three minimal natural scenarios can be formulated, which provides a connection between the QCD-axions and mediators at the GUT/string scales. We find that the PQ symmetry breaking scale could be about $10^{15}$ GeV, higher than the classical QCD dark matter axion window but possible if the anthropic window is considered. We also propose an experiment to probe such scenarios. If the dark matter axion is discovered, they might suggest that we live in an atypical Hubble volume.
\end{abstract}

\maketitle

{\bf Introduction:} The Peccei-Quinn (PQ) mechanism provides a
 well-motivated and elegant solution to the strong CP problem \cite{Peccei:1977hh,Peccei:1977ur, Weinberg:1977ma, Wilczek:1977pj,Vysotsky:1978dc}. In it, a global $U(1)$ PQ symmetry was broken both spontaneously and explicitly in the early Universe, which produced very cold cosmological relic density of a pseudo-goldstone boson, the axion. The creation mechanism is now known as the misalignment mechanism \cite{Preskill:1982cy, Abbott:1982af, Dine:1982ah,Sikivie:1982qv,Ipser:1983mw,Sikivie:2006ni}. There are many promising axion models such as the KSVZ axions and DFSZ axions etc. \cite{Kim:1979if, Shifman:1979if,Zhitnitsky:1980tq, Dine:1981rt,Berezhiani:1990jj,Berezhiani:1989fu,MartinCamalich:2020dfe}. The created cosmic axions can constitute most of dark matter if the classical QCD dark matter axion window $f_a\sim {\cal O}(10^{11})$~GeV, $m_a\sim {\cal O}(10^{-5})$eV is satisfied, which could be experimentally tested \cite{Sikivie:1983ip,Sikivie:1985yu}, although the other dark matter axion windows are possible \cite{Hertzberg:2008wr}.

Axion-like particles (ALPs) are generally created from the extra dimension compactifications \cite{Svrcek:2006yi} which is often inevitable for a UV theory such as the string theory. Due to the uncertainties of the instanton effects on the compactified dimensions, ALPs have a relaxed $f_a-m_a$ correlation, and thus have a much larger window to serve as the dark matter particles. The general ALPs, however, may not be connected to the strong CP problem.

It is interesting to consider a scenario which has the minimal UV-generated effective operators that couple to the QCD sector. In this paper, we consider a minimal complex singlet scalar extension of the Standard Model (SM) in which the SM fermions and the singlet are carrying the U(1) PQ symmetry charges. Three possible constructions with effective coupling operators generated from the GUT/string scale are considered. The phenomenological consequences and a possible laboratory probe is discussed.

{\bf Theoretical formulation:} A UV theory typically can be written into some effective operators in a lower energy scales which sometimes is necessary to study the UV theory without abundant phenomenological data. Here, we propose a minimal extension in which the SM fields are charged under a high scale $U(1)$ PQ symmetry and a complex scalar $S$ breaks it spontaneously. There are three natural realizations of the effective operators:

{\itshape Scenario I:} the PQ charges are assigned as follows
\be
\begin{tabular}{ccc}
$Q_i,~L_i,~U_i^c,~D_i^c, E^c_i$ & $:$ & 1 \\
$H_u$ & $:$ & -2 \\
$S$ & $:$ & -4
\end{tabular}
\ee
where the subscripts $i=1,2,3$ are the flavor index. $H_u$ is the SU(2) doublet that break the electroweak
gauge symmetry. The induced effective Lagrangian are
\bea
{\cal L} &=& -y_{ij}^u Q_i U_j^c H_u - y_{ij}^d \frac{S}{M_*}Q_i D_j^c\tilde{H}_u, \\
 & & - y_{ij}^e \frac{S}{M_*} L_i E_j^c \tilde{H}_u ~~, \nonumber
\label{eq:lag1}
\eea
where $\tilde{H}_u = i\sigma_2 H_u^*$. The Yukawa terms of the SM down-type quarks and the leptons are formulated by the UV effective operators. The phase factor of $S$ is the axion after the $S$ receives a vacuum expectation value $\vev{S}$. In case the $M_*$ represents the physics at the string or the reduced Planck scale, the effective axion PQ scale will be $f_a\sim 10^{15-16}$ GeV. Note that this scale is about 4 orders of magnitude larger than the classical window of the QCD dark matter axions, therefore the PQ symmetry breaking could be happened before the inflation and during which an anthropic hubble volume was selected \cite{Hertzberg:2008wr}.

The effective operators can be generated by integrating the heavy vector-like paricles as well.
For example, we introduce the vector-like fields $({\xi}_i, {\xi}_i^c)$ with a PQ charges $(3,~-3)$ at $10^{12}$~GeV,
and $y^d_{ij}$ terms can be generated  from integrating of the $({\xi}_i, {\xi}_i^c)$. The relevant Lagrangian is
\be
{\cal L}_{{\xi}} = -y^{\prime d}_{ij} Q_i {\xi}_j^c \tilde{H}_u - y'_{lk} S D_l^c {\xi}_k +M^{{\xi}}_{jk} {\xi}^c_j {\xi}_k, \nonumber
\ee
which yields the effective operators
\be
{\cal L}_{\rm eff.} = \frac{y_{ij}^{\prime d} y'_{kl}}{M^{{\xi}}_{jk}} Q_i D_l^c \tilde{H}_u S~,~
\ee
where we have $y_{i1}^d \equiv  -y_{i1}'^d y'$ and $M_* \rightarrow M^{{\xi}}$. Since $M^{{\xi}}$ can be lower
than the GUT scale around $10^{12}$ GeV, this scenario is consistent with the classical QCD dark matter axion window.
The CKM quark mixing
matrix can be obtained by assigning proper $y_{ij}^u$ to the up-type quark sector.
The discussions for all the other effective operators are similar.
However, such kind of scenarios is not minimal.

{\itshape Scenario II:} the PQ charges assume the following assignment:
\be
\begin{tabular}{ccc}
$Q_i,~L_i,~U_i^c,~D_1^c$ & $:$ & 1 \\
$D_2^c,~ D_3^c,~ E_i^c$ & $:$ & -3 \\
$H_u$ & $:$ & -2 \\
$S$ & $:$ & -4
\end{tabular}
\ee
This scenario differs from the scenario I in the $D_2,~D_3$ and the $E$ fields:
\bea
\label{eq:lag2}
{\cal L} &=& -y_{ij}^u Q_i U_j^c H_u - y_{ik}^d Q_i D_k^c\tilde{H}_u, \\
 & & - y_{ij}^e L_i E_j^c \tilde{H}_u + y_{i1}^d \frac{S}{M_*} Q_i D_1^c \tilde{H}_u\nonumber
\eea
where $i,j=1,2,3$ and $k=2,3$. The effective operators involve the light 1st-generation $d-$quark. Because $m_d\sim 4.7$ MeV, we have $\vev{S}\sim 10^{12}$ GeV if $M_*$ is at the string scale. Also, it has $\frac{E}{N}=\frac{2}{3}$.

{\itshape Scenario III:} the particles carry PQ charges as below
\be
\begin{tabular}{ccc}
$Q_i,~L_i,~U_2^c,~U_3^c$ & $:$ & 1 \\
$U_1^c,~ D_i^c,~ E_i^c$ & $:$ & -3 \\
$H_u$ & $:$ & -2 \\
$S$ & $:$ & -4
\end{tabular}
\ee
where the UV effective operators are assigned to the up-type quark sector therefore
\bea
\label{eq:lag3}
{\cal L} &=& -y_{ik}^u Q_i U_k^c H_u - y_{i1}^u \frac{S^*}{M_*}Q_i U_1^c H_u, \\
 & & - y_{ij}^d Q_i D_j^c \tilde{H}_u + y_{ij}^e L_i E_j^c \tilde{H}_u \nonumber
\eea
where $i,j=1,2,3$ and $k=2,3$. With $m_u\approx 2.2$MeV, $\vev{S}\sim 10^{12}$~GeV, it is compatible with $M_*$ of the string scale. This scenario is similar to the scenario II, and has $\frac{E}{N}=\frac{8}{3}$.

{\bf Baryon number and flavor violation:} The operators with net baryon and lepton numbers in Scenario I can be written as
\bea
~\frac{\epsilon^{\eta\rho}\epsilon^{\delta \sigma} Q_\eta Q_\rho Q_\delta L_{\sigma} S}{M^3_*}
&\sim & \frac{\vev{S}}{M_*} \cdot
\frac{\epsilon^{\eta\rho}\epsilon^{\delta \sigma}}{M_*^{2}} Q_\eta Q_\rho Q_\delta L_{\sigma},
\nonumber \\
~\frac{U^cU^cD^c E^c S^*}{M^3_*}
&\sim &
\frac{\vev{S}}{M_*} \cdot
\frac{1}{M_*^{2}}U^cU^cD^c E^c,
\eea
where the $B-L$ number is still preserved, and the anticommuting $\{\eta, \rho; \delta, \sigma \}=\{1,2\}$ refer
to $SU(2)_L$ doublet indices. After U(1) symmetry breaking, these effective operators become similar to the well known GUT-scale mediated baryon number violating operators but with a fore-factor $\vev{S}/M_*$. The resulting contribution to the proton decay rate is expected to scale by ${\cal O}(\vev{S}^2/M^2_*)$ in comparison to that from 4-fermion GUT induced operators. Similarly, we can study the Scenarios II and III.

Also note the quark PQ charges are not flavor universal, the effective operators with $S$ field can violate quark flavor, as seen in Eq.~\ref{eq:lag2} and ~\ref{eq:lag3} for Scenario II and III. Flavor off-diagonal quark Yukawa couplings are generated as $\left< S \right>/M_*\sim {\cal O}(10^{-4})$ for $\vev{S}\sim 10^{12}$ GeV and $M_*$ at the string scale. Such couplings do not contribute to FCNC since one Higgs doublet is present at low scale in the model~\cite{Glashow:1970gm}, and similarly flavor off-diagonal axion-quark couplings obtained via Im${H_u}$ are rotated away. However, the dim-5 effective operator can still generate flavor changing axion-quark couplings:
\be
g_{aqq'}=\sum_{i=1,2,3}\frac{\vev{H_u^0}}{M_*}{\cal V}_{qi}^\dagger y^{q,{\rm dim-5}}_{i1}{\cal U}_{1q'},
\ee
where $q=d$ for Scenario II and $u$ for Scenario III respectively, and $q$ iterates over the three generations on quark mass eigenstates. ${\cal U, V}$ are the bi-unitary rotation matrices that diagonalize the quark mass matrix, and $y^{q,{\rm dim-5}}$ only includes the dim-5 operator's couplings.
$g_{aqq'}$ are suppressed by a factor of $\vev{S}/M_*$ compared to the flavor-diagonal axion-quark couplings. Thus our flavor-changing axion-quark couplings correspond to a rather high scale $f_{aqq'}=C_{qq'}^{-1} M_*$, where $C_{qq'}$ is a sub-unity quantity due to flavor mixing. For string scale $M_*$ these off-diagonal couplings are allowed by current kaon decay~\cite{Fantechi:2014hqa,Beckford:2017gsf} and cosmological~\cite{DEramo:2021usm} constraints. In case of a lower scale $M_*$ close to that of $\vev{S}$, flavor-changing meson decays could give significant constraints for $f_{aqq'}\sim C_{qq'}^{-1}\vev{S}$ \cite{Bjorkeroth:2018dzu}.

{\bf Cosmological implications:} In the following discussions, we will consider the major phenomenological implications when $f_a\sim 10^{15}$~GeV.
The minimal models lead to a possibility of $f_a\sim 10^{15}$~GeV which will produce too much dark matter if the PQ symmetry breaking happened after inflation. However, if the PQ symmetry happened before the early time of inflation an anthropic window is possible. The major constraint is then the inflation Hubble scale $H_I$ and the later generated isocurvature perturbations.

The axion field evolves in an expanding flat FRW universe as
\be
\partial^2_t a+3H\partial_t a-{1\over R^2}\nabla ^2a+\partial_a V(a)=0~~,
\ee
where $R$ is the scale factor, $H=\dot R/R$ is the Hubble parameter, and $V(a)$ is the potential of the axion field. The potential term is written as
\be
V(a)\approx f_a^2m_a^2(T)\left[1-{\rm cos}({a\over f_a})\right]~~.
\ee
If $T>\Lambda_Q\sim 200$MeV, the mass is temperature dependent
\be
m_a(T)\propto m_0b\left({\Lambda_Q\over T}\right)^4~~,
\ee
where $b\sim{\cal O}(10^{-2})$ and when $T\lesssim \Lambda_Q$, the axion mass $m_0$ is almost a constant.

When $H$ was very large, the axion potential $V$ could be neglected so the initial misalignment angle $\theta=a/f_a$ was frozen. The axions started to oscillate at $3H\approx m_a(T_0)$ and the resulting energy density at temperature $T_0$ was
\be
\rho_a\sim {1\over2}m_a(T_0)^2\vev{a_0^2}~~.
\ee
The average value of the field square over the universe is determined by the initial misalignment angle $\theta_0$ and its standard deviation $\sigma_{\theta}$
\be
\vev{a_0^2}=(\theta_0^2+\sigma_{\theta}^2)f_a^2~~,
\ee
Notice that the axion energy density was transferred from the QCD sector so the total energy density was conserved
\bea
\delta \rho_{total}=m_a\delta n_a+m_i\delta n_i+4\rho_{rad}{\delta T\over T}=0.
\eea
The fluctuation of number density $n_i$ to entropy density is
\bea
\delta S_i={\delta (n_i/s)\over n_i/s}=\delta n_i/n_i-3\delta T/T.
\eea
The axion energy density was small comparing to the total energy density initially, and thus
\bea
\delta S_a & \approx &\delta n_a/n_a= {\theta^2-\vev{\theta^2}\over \vev{\theta^2}}~.~\,
\eea
Assuming $\delta \theta=\theta-\vev{\theta}$ is Gaussian, we obtain
\bea
~\vev{\delta S_a^2} & = &{2\sigma^2_{\theta}(2\theta_0^2+\sigma _{\theta}^2)\over(\theta_0^2+\sigma_{\theta}^2)^2}~~.
\eea

As the $U(1)$ PQ symmetry breaking happened before inflation, the standard deviation is at
the order of
\be
\sigma_\theta\sim {H_I\over 2\pi f_a}~~,
\ee
due to the Gibbons-Hawking temperature at inflation.

Denoting the energy density of the particle specie $i$ to the photon number density as $\xi_i (T)=\rho_i(T)/ n_{\gamma}(T)$, the isocurvature perturbation is
\bea
{\vev{\left({\delta T\over T}\right)_{\rm iso}^2}}\sim \left({\xi_a\over 3\xi_{\rm matter}}\right)^2\vev{\delta S_a^2}\lesssim {\cal O} (10^{-11}).
\eea
Considering low scale inflation scenarios such as $H_I\ll 10^{10}$~GeV, the PQ symmetry breaking scale is then constrained as
\be
f_a\gtrsim 10^{14}{\rm ~GeV}~~,
\ee
if the axions constitute a majority part of dark matter which should be consistent to our minimal model.

{\bf Experimental probe:} With $f_a\sim 10^{15}$~GeV the axion has a mass about $10^{-8}$~eV. The energy per axion is very small. In addition, the couplings to the SM particles are suppressed by $f_a$, which makes experimental detection difficult. It is, however, possible to probe the scenario using the hyperfine splitting of atoms \cite{Yang:2019xdz}. The hydrogen $1S$ state is split into four energy eigenstates, which, under a very weak external magnetic field, can be written as a spin-0 singlet $\bra{0,0}$ and three spin-1 triplet states $\bra{1, m_j}$ where $j, m$ are the quantum number for the total spin and its $z$-component. The energy gap $\Delta E$ between $\bra{1, 0}$, and $\bra{1, 1}$ is narrow if the external magnetic field is sufficiently small
\bea
{\Delta E}/{{\rm eV}}&=& 2.95\times 10^{-6}+2.58\times 10^{-5}(B/T) \nonumber\\
&-&2.95\times 10^{-6}\sqrt{1+76.48(B/T)^2} ~~.
\eea
For example, when $B\approx 0.001$T, $\Delta E \approx 2.5\times 10^{-8}$~eV which is close to the axion mass in our scenario. When the energy gap is adjusted to match the axion mass, $\bra{1,0}\to \bra{1,1}$ transition will be induced at a rate
\bea
&~&~NR=N{\pi \over f_a^2m_a}\left({v\over \delta v}\right)^2\rho_{\rm DM}\\ \nonumber
&~&=N\cdot 3\times 10^{-22}\left({ v\over \delta v}\right)^2{ \left({m_a\over {\rm 10^{-5}eV}}\right)^{-1}\left({f_a\over 10^{11}{\rm GeV}}\right)^{-2}} {\rm s}^{-1},
\eea
where $v\sim 3\times 10^{-3}$ is the dark matter's velocity relative to the laboratory. Typically $\delta v\sim 10^{-3}$ for the velocity distribution in our Galactic halo, $\rho_{\rm DM}\approx1$~GeVcm$^{-3}$ is the local halo density, and $N$ is the amount of atom targets. For $f_a\sim 10^{15}$~GeV, $m_a\sim 10^{-8}$ eV and $N\sim 10^2$ moles, the event rate is 16.2 per second. High $\bra{1,0}$ occupation number atoms can prepared by Stern-Gerlach apparatus to filter $m\neq 0$ states. The axion-induced transition flips the state to $\bra{1,1}$ which could be countable using the Stern-Gerlach apparatus again, and the spontaneous $\bra{1,0}\rightarrow \bra{0,0}$ decay does not cause contamination as $m$ remains unchanged. In addition, the spontaneously $\bra{1,0}\rightarrow \bra{1,-1}$ decay could be distinguished as the resulted atomic diffraction is opposite in the Stern-Gerlach apparatus. Assuming the system is cooled down to a temperature that thermal noise is negligible and the atomic spacings are sufficient large to avoid collision induced transitions, the sensitivity with one year scanning is
\bea
f_a\leq 8.1\times 10^{12}{\rm GeV}\sqrt{{{\rm GHz}\over\Delta f}}\sqrt{{N\over {\rm mole}}}~~.
\eea
Using $N$=$10^2$ moles, the covered mass range $\Delta f\sim 0.1$~MHz $\sim 10^{-8}$~eV, and the $U(1)$ PQ symmetry
breaking scale can be tested up to $f_a\sim 8.1\times 10^{15}$~GeV.

In additon to the induced atomic transitions, we also note our model with $f_a\sim 10^{15}$~GeV falls
in the optimal detection range of a LC curcuit experimental desgin~\cite{Sikivie:2013laa}
using haloscope or accelerator magnets.

{\bf Summary:} Considering that the UV-induced effective operators generate the necessary couplings
before the PQ symmetry breaking, and any low-scale couplings to the SM are restricted to the Yukawa sector,
we proposed three minimal natural axion models. They might provide a connection between the QCD-axions
and mediators around the GUT/string scales. We found that the PQ symmetry breaking scale could
be about $10^{15}$ GeV, higher than the classical QCD dark matter axion window but possible
when the anthropic window is considered. We also proposed the experimet to test these scenarios.
And if the dark matter axion is discovered, it seems that we might live in an atypical Hubble volume.

\medskip

{\bf Acknowledgments:}
We would like to thank Pierre Sikivie, Nick Houston and Xin Zhang for helpful discussions. Y.G. is supported by Grant No. Y95461A0U2, Institute of High Energy Physics, CAS, by No. 12150010 supported by the National Natural Science Foundation of China, and in part by the Ministry of Science and Technology of China (2020YFC2201601). T.L. is supported by the National Key Research and Development Program of China Grant No. 2020YFC2201504, by the Projects No. 11875062, No. 11947302, and No. 12047503 supported by the National Natural Science Foundation of China, as well as by the Key Research Program of the Chinese Academy of Sciences, Grant NO. XDPB15. Q.Y. is supported by the National Natural Science Foundation of China under Grant No. 11875148, No. 12150010 and is funded in part by the Gordon and Betty Moore Foundation through Grant GBMF6210.


\bibliography{refs}

\end{document}